\documentclass[twocolumn,prl,superscriptaddress,showpacs]{revtex4-2}
\newcommand{\pagenumbaa}{1}
\usepackage{amsmath}
\usepackage{graphicx}
\usepackage{setspace}
\usepackage{units}
\usepackage{color}
\usepackage{hyperref}
\usepackage[dvipsnames]{xcolor}

\begin{document}
\title{Isotropic and Anisotropic g-factor Corrections in GaAs Quantum Dots}

\author{Leon C. Camenzind}
\thanks{These authors contributed equally.}
\affiliation{Department of Physics, University of Basel, Basel 4056, Switzerland}

\author{Simon Svab}
\thanks{These authors contributed equally.}
\affiliation{Department of Physics, University of Basel, Basel 4056, Switzerland}

\author{Peter Stano}
\affiliation{Center for Emergent Matter Science, RIKEN, Saitama 351-0198, Japan}
\affiliation{Institute of Physics, Slovak Academy of Sciences, 845 11 Bratislava, Slovakia}

\author{Liuqi Yu}
\altaffiliation{Present address: Laboratory for Physical Sciences, 8050 Greenmead Drive, College Park, MD 20740, USA.}
\affiliation{Department of Physics, University of Basel, Basel 4056, Switzerland}

\author{Jeramy D. Zimmerman}
\altaffiliation{Present address: Physics Department, Colorado School of Mines, Golden, CO, 80401, USA.}
\affiliation{Materials Department, University of California, Santa Barbara 93106, USA}

\author{Arthur C. Gossard}
\affiliation{Materials Department, University of California, Santa Barbara 93106, USA}

\author{Daniel Loss}
\affiliation{Department of Physics, University of Basel, Basel 4056, Switzerland}
\affiliation{Center for Emergent Matter Science, RIKEN, Saitama 351-0198, Japan}

\author{Dominik M. Zumb\"uhl}
\affiliation{Department of Physics, University of Basel, Basel 4056, Switzerland}

\date{21 / 10 / 2020}
\setcounter{page}{\pagenumbaa}
\thispagestyle{plain}

\begin{abstract}
We experimentally determine isotropic and anisotropic g-factor corrections in lateral GaAs single-electron quantum dots. We extract the Zeeman splitting by measuring the tunnel rates into the individual spin states of an empty quantum dot for an in-plane magnetic field with various strengths and directions. We quantify the Zeeman energy and find a linear dependence on the magnetic field strength which allows us to extract the g-factor. 
The measured g-factor is understood in terms of spin-orbit interaction induced isotropic and anisotropic corrections to the GaAs bulk g-factor. Because this implies a dependence of the spin splitting on the magnetic field direction, these findings are of  significance for spin qubits in GaAs quantum dots. 
\end{abstract}
\maketitle

\makeatletter{\renewcommand*{\@makefnmark}{}
\footnotetext{\textsuperscript{$*$}\textit{These authors contributed equally to this work}.}\makeatother}

\makeatletter{\renewcommand*{\@makefnmark}{}
\footnotetext{\textsuperscript{$\dagger$}\textit{Present address: Laboratory for Physical Sciences, 8050 Greenmead Drive, College Park, MD 20740, USA.}}\makeatother}

\makeatletter{\renewcommand*{\@makefnmark}{}
\footnotetext{\textsuperscript{$\dagger\dagger$}\textit{Present address: Physics Department, Colorado School of Mines, Golden, CO, 80401, USA.}}\makeatother}


Spins in semiconductor quantum dots are among the most promising candidates for the realization of a scalable quantum bit (qubit) \cite{Loss1998,Kloeffel2013}. For such spin qubits, the qubit energy is the Zeeman energy $\Delta = g \mu_{\rm B} B$, where $\mu_{\rm B}$ is the Bohr magneton, $B$ is the magnetic field and $g$ is the g-factor. Hence, a detailed understanding of the g-factor is an important ingredient for the addressability of spin qubits. In multi-qubit devices, local g-factor differences between the individual qubits allow to address them selectively, and can also be utilized for the realization of quantum logic gates \cite{Veldhorst2015, Jones2018}.

In addition to the addressability, the g-factor can also impact the coherence of spin qubits.
While the spin of an electron in GaAs is strongly influenced by the host nuclear spins through the hyperfine interaction \cite{Merkulov2002, Khaetskii2002, Petta2005, Koppens2008, Cywinski2009, Bluhm2011}, the resulting magnetic noise is slow \cite{Delbecq2016}, allowing for effective countermeasures: The short dephasing time of an unprotected spin, of order 10 ns \cite{Petta2005, Koppens2008}, has been extended to order $\mu$s by postprocessing \cite{Nakajima2019} or active compensation \cite{Shulman2014, Nakajima2020} and to order ms by dynamical decoupling \cite{Malinowski2016}. At such long time scales, additional decoherence sources need to be considered, and fluctuations in the g-factor are one of them.
These fluctuations originate from charge noise as $g$ is sensitive to the local electric field \cite{Veldhorst2014,Ferdous2018}. In addition, this type of decoherence will be a major issue in group-IV semiconductors with little or no nuclear spins, such as silicon \cite{Tanttu2018} and Si/SiGe heterostructures \cite{Yoneda2018}.

In semiconductors, g-factor corrections arise from the spin-orbit interaction (SOI) \cite{Winkler2003,Stano2018gfactor}, and spatial variations of $g$ might occur due to local electric fields which modulate this interaction \cite{Tanttu2018,Ferdous2018,Stano2018gfactor}. Recently, measurements in a SiMOS spin qubit \cite{Tanttu2018} showed that these corrections are small for electrons in silicon due to comparably weak SOI. For holes, experiments in silicon MOSFETs \cite{Voisin2016, Crippa2018}, in a GaAs heterostructure \cite{Wang2016}, and in a silicon-germanium core-shell nanowire \cite{Brauns2016} show that these corrections are more pronounced due to stronger SOI \cite{Kloeffel2013}.

Here, we present an experiment where we can separate the isotropic and anisotropic g-factor corrections in two GaAs single-electron spin qubits with slightly different wafer properties. In a recent model by \textit{Stano et al.} \cite{Stano2018gfactor}, the Rashba SOI, together with a bulk structure SOI term at finite magnetic field, lead to isotropic corrections, while the Dresselhaus SOI is giving rise to an anisotropic correction. The experiment is in good agreement with these predictions and thus provides clear evidence for a profound, detailed model of the g-factor corrections, serving as the key characteristic of a GaAs quantum dot spin qubit. Previous experiments in this material did not identify the g-factor anisotropy \cite{Michal2018, Zumbuhl2004}.

\begin{figure}
	\centering
		\includegraphics[width=0.95\columnwidth]{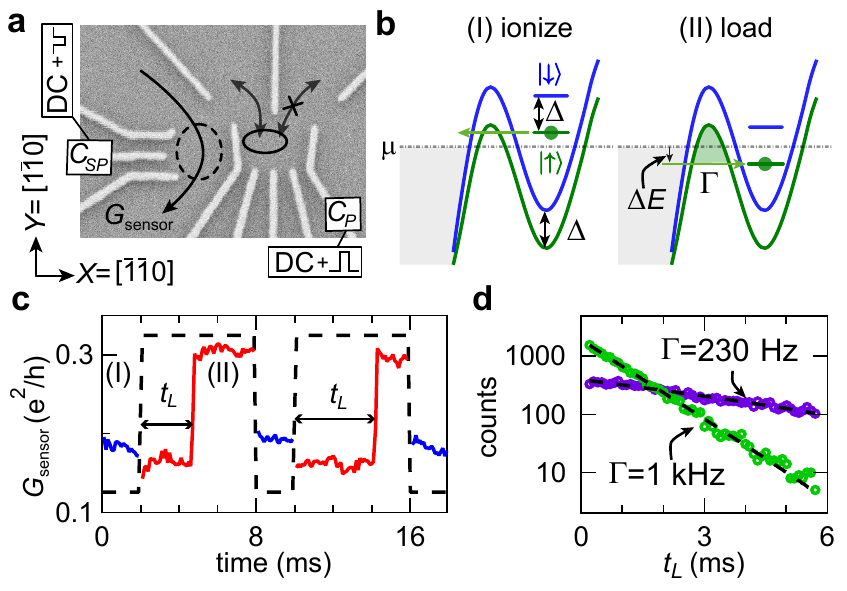} \vspace{-3mm}
	\caption{(a) Electron micrograph of a cofabricated device with dot position (solid ellipse) and sensor dot (dashed ellipse). The sensor conductance $G_{\text{sensor}}$ reads the real-time charge state of the dot.
	(b) Two-step pulses (I) ionize and (II) load applied on the dot gate $C_P$, used to measure the tunneling rate $\Gamma$ into the empty dot for detuning $\Delta E$ from the reservoir chemical potential $\mu$. The sensor plunger $C_{SP}$ is compensated to maintain read-out sensitivity. A magnetic field splits the dot states $\left|\uparrow\right\rangle$, $\left|\downarrow\right\rangle$ as well as the conduction band (blue and green) by the Zeeman energy $\Delta$.
	(c) Sensor conductance $G_{\text{sensor}}$ for two cycles (dashed pulses). Low (high) $G_{\text{sensor}}$ indicates an empty (occupied) dot, respectively. The ionization rate during (I) is faster than the sensor bandwidth. The electron loading times $t_L$, appearing as clear steps (red traces), are histogrammed to extract the tunnel rate $\Gamma$ via exponential fit, shown in (d) for two examples, with typical error bars $\pm 10$~Hz. }
	\label{fig:gfactor:Fig1}\vspace{-4mm}
\end{figure}

The experiment was performed on two separate quantum dots, each in the single-electron regime, with adjacent quantum dot charge sensor, all defined with depletion gates, layout shown in Fig.~\ref{fig:gfactor:Fig1}a, on two slightly different GaAs 2D electron gases (2DEGs), see Sec.~1 in Ref.~\cite{som} for details. The crystal axes as labeled were carefully tracked from the wafer flats. Here, the quantum dot is tunnel coupled only to the left reservoir. The sensor conductance reads the charge state \cite{Field1993,Barthel2010}, here with a bandwidth of $\sim$30\,kHz. The device is mounted on a piezo rotator stage (Attocube ANRv51), allowing magnetic fields up to 14\,T to be applied in an arbitrary in-plane direction. The misalignment is less than 2$^{\circ}$ and thus here negligible \cite{Camenzind2017}. Measurements are carried out in a dilution refrigerator at an electron temperature of 200 mK.

To calculate the Zeeman energy $\Delta$, it is necessary to convert changes of the voltage applied to the plunger gate $C_P$ to changes in energy of the quantum dot levels (see Fig.~\ref{fig:gfactor:Fig1}b). We calibrate the lever arm of $C_P$ by probing the Fermi-Dirac distribution of the reservoir at an increased temperature ($\sim$600\,mK) where we assume that the electronic temperature is the same as the temperature of the mixing chamber \cite{Maradan2014}. We checked that the lever arm shows no significant dependence on the strength or direction of the external magnetic field. Nevertheless, the lever arm ends up giving the dominant contribution to the accuracy of the extracted Zeeman splitting and g-factors, as will be discussed later. The right barrier directly next to $C_P$ is opaque, and the left reservoir barrier is relatively far from $C_P$, such that voltage changes on this plunger-gate do not affect the tunneling barrier by much.

Because the g-factor corrections depend on the shape of the quantum dot, we performed a recently developed spectroscopy of the quantum dot orbitals with in-plane magnetic fields \cite{Camenzind2018}. For device 1, we found orbital energies that suggest a slightly ellipsoidal, disc shaped quantum dot which is elongated along the x-axis of the device as depicted in Fig.~\ref{fig:gfactor:Fig1}a. For device 2, we found a more asymmetric dot shape than in the first device. We note that for this difference in dot shapes, the theory predicts an immeasurable small difference in the g-factor (see Sec.~II in Ref.~\cite{som}).

We obtain $g$ by measuring the tunnel rate $\Gamma$ into the spin states of an empty quantum dot, taking advantage of the increase in $\Gamma$ when both spin states are energetically available. From these rates we extract $\Delta$, and from the dependence of $\Delta$ on the magnetic field strength we fit $g$. We measure $\Gamma$ by applying a two-step pulse to plunger gate $C_P$ (see Fig.~\ref{fig:gfactor:Fig1}a), repeatedly ionizing and loading the quantum dot as shown by the energy diagrams in Fig.~\ref{fig:gfactor:Fig1}b: to ionize, the energy level of the charged quantum dot is pulsed above the chemical potential $\mu$ of the reservoir such that an electron will tunnel into the empty states of the reservoir and thermalize.
We chose this ionization pulse such that the ionization efficiency is close to unity.
To load, we pulse the empty quantum dot to an energy detuning $\Delta E$ below $\mu$. At this energy, filled states are available in the reservoir and an electron can elastically tunnel through the barrier into the quantum dot. The time constant of this probabilistic tunnel process is given by $\Gamma$.

We obtain $\Gamma$ by monitoring the charge sensor conductance $G_{\text{sensor}}$ and extract the times of these loading events $t_L$ as shown in Fig~\ref{fig:gfactor:Fig1}c: the tunneling of an electron leads to a change of the charge state from empty to loaded, which results in an observable switch to an higher $G_{\text{sensor}}$. We cycle through this pulse scheme between 2k and 20k times and extract $\Gamma$ by fitting an exponential function to a histogram of $t_L$ (see Fig~\ref{fig:gfactor:Fig1}d). When changing the pulse amplitudes, we obtain $\Gamma$ as a function of the detuning $\Delta E$.

Two important comments about the experiment: first, to stay in the sweet spot of the sensor during the pulse sequence, we have to compensate the crosstalk between the pulses applied to $C_P$ and the sensor quantum dot by applying pulses of opposite polarity to the sensor plunger gate $C_{SP}$ (see Fig.~\ref{fig:gfactor:Fig1}a) \cite{Biesinger2015}.
Second, we divide the total number of pulse cycles into segments in order to mitigate drift-related effects: In every segment, 100 pulses are applied at each selected detuning $\Delta E$ before an automated feed-back loop is used to compensate for time-dependent drifts of the quantum dot levels by retrieving the position of $\Delta E = 0$ \cite{Amasha2008a}. We exclude hysteresis effects by selecting the sequence of detunings $\Delta E$ to which we pulse randomly for each round.

\begin{figure}[t]
	\centering
		\includegraphics[width=1\columnwidth]{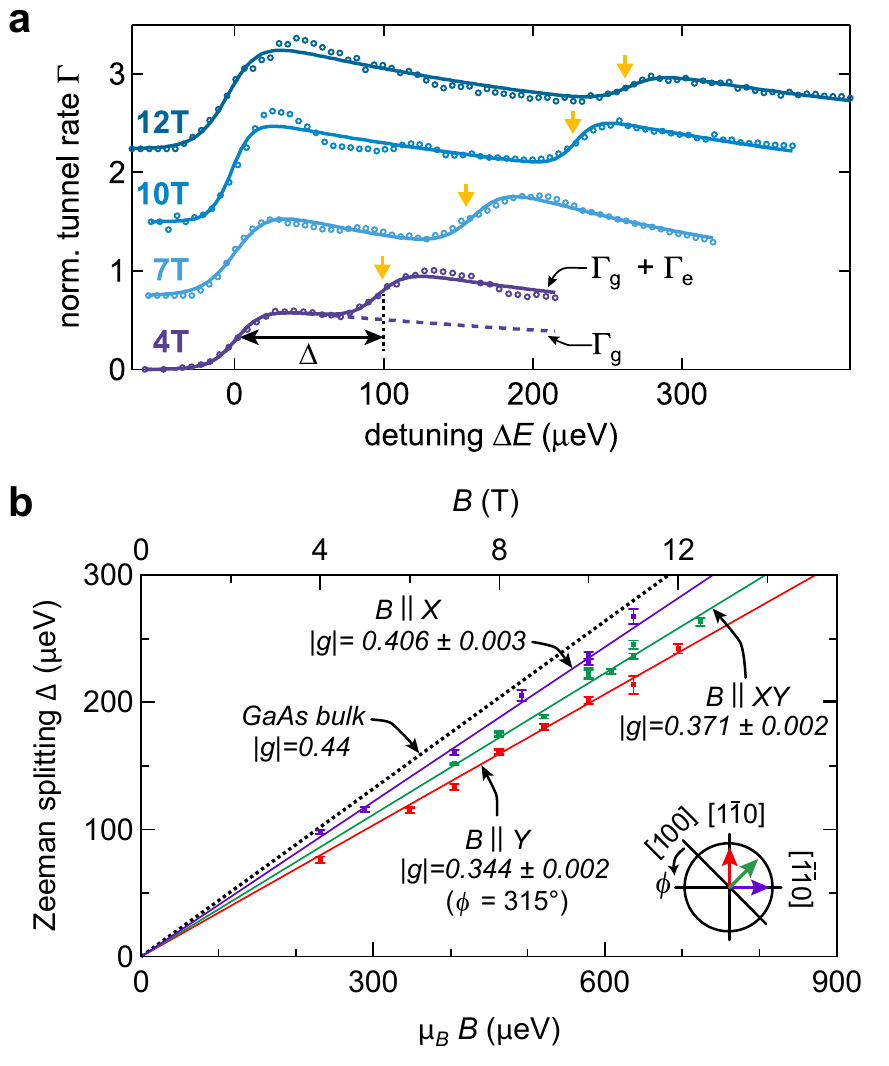} 
	\caption{(a) Examples of the normalized tunnel rate $\Gamma$ into the empty quantum dot for different detunings $\Delta E$ and magnetic field strengths. Each trace exhibits two resonances, identified as the two spin states due to their behavior in magnetic field (yellow arrows). The fits shown here are according to a phenomenological model described in Sec.~3 of Ref.~\cite{som}.
 In the trace taken at 4\,T, the dashed line shows $\Gamma_g (\Delta E)$, the contribution of the spin ground state to the total tunnel rate, and $\Delta$ indicates the Zeeman splitting. (b) Zeeman splittings $\Delta$ in device 1, measured for different magnetic field strengths $B$ and directions as indicated by the labels. The error bars reflect the statistical uncertainty from the fits. The slope is the absolute value of the g-factor $\vert g\vert =\Delta/(\mu_{\rm{B}} \vert B\vert)$ and differs from the GaAs bulk g-factor due to spin-orbit interaction induced corrections. A distinct g-factor anisotropy is observed in the data. The inset shows the direction of the applied magnetic fields with respect to the crystal axes.}
	\label{fig:gfactor:Fig2} \vspace{-4mm}
\end{figure}

In Fig.~\ref{fig:gfactor:Fig2}a we show data of $\Gamma (\Delta E)$ for increasing magnetic fields up to 12~T. Due to orbital effects of the in-plane magnetic field \cite{Camenzind2018}, the tunnel barriers have to be readjusted for each field configuration in order to keep the tunnel rates at a couple of hundred Hz (see Sec.~4 in Ref.~\cite{som}). As a consequence, the magnitudes of $\Gamma(\Delta E)$ for the different traces are not comparable and were therefore normalized in Fig.~\ref{fig:gfactor:Fig2}a. As the dot ground state is pulled below the reservoir and $\Delta E$ starts to increase from zero, electrons may start to tunnel onto the dot, leading to the rising flank as seen in Fig. \ref{fig:gfactor:Fig2}a for $\Delta E\gtrsim 0$. The observed broadening is given by the reservoir temperature. As the dot level is pulled further below the reservoir, eventually also the excited spin state becomes available, thus increasing the tunnel rate above the ground state rate, as indicated by the yellow arrow. The separation of the two steps is thus identified as the Zeeman splitting $\Delta$, and grows with magnetic field, as seen on Fig. \ref{fig:gfactor:Fig2}a. The observed exponential suppression of $\Gamma$ with increasing $\Delta E$ is attributed to an effective increase of the tunnel barrier potential experienced by the electrons when the gate voltage is increased \cite{MacLean2007,Amasha2008b,Stano2010,Simmons2011}.

Next, we look at the magnetic field strength and direction dependence of the extracted $\Delta$ to investigate the behavior of the g-factor. We take such data for magnetic fields applied in a range of directions between the crystallographic axes $[\bar{1}\bar{1}0]$ ($X$) and $[1\bar{1}0]$ ($Y$) (see Fig. \ref{fig:gfactor:Fig1}a). The measured Zeeman splittings $\Delta$ for device 1 are plotted in Fig.~\ref{fig:gfactor:Fig2}b (see Sec.~5 in Ref.~\cite{som} for device 2). We find a linear dependence for all directions, which indicates that the g-factor is independent of the strength of the magnetic field. Accordingly, we use a linear fit (without offset) on these data sets to obtain $\vert g \vert$, the absolute value of $g$. The statistical uncertainty obtained from the fits is in the range of one percent relative error. Also, it was not possible to obtain a reliable $\Delta$ at some specific $B$ values and directions due to vanishing excited spin tunneling \cite{Amasha2008a, Yamagishi2014} and/or due to measurement artifacts such as reservoir resonances.

Strikingly, our data indicates that $g$ depends on the magnetic-field direction. For device 1, the g-factor is maximal for a field along $X$ with $\vert g \vert \approx 0.406$, and minimal along $Y$ where $\vert g \vert \approx 0.344$. This difference is well above the statistical error bar, and similar in device 2 (see Sec.~5 in Ref.~\cite{som}). This is in good qualitative agreement with the theory in Ref.~\cite{Stano2018gfactor}. In that model, there are numerous terms giving corrections to the bulk g-factor. These can be separated into an isotropic and an anisotropic part, such that
\begin{equation}
	g = g_{\text{bulk}} + \delta g_{i} + \delta g_{a}\cos\left({2\phi + \pi/2}\right),
	\label{eq:gCorrections}
\end{equation}
where $g_{\text{bulk}} =-0.44$ is the GaAs bulk g-factor and $\phi$ defines the in-plane angle with respect to the main crystal axis $[100]$ (see inset in Fig.~\ref{fig:gfactor:Fig2}b). Here, terms with higher-order angle dependence are small and are neglected. We extract $\delta g_i$ and $\delta g_a$ experimentally, and the quantification of these two parameters for our quantum dot is the main result of this article. For most of the relevant terms, the magnitudes of the g-factor corrections depend primarily on $\lambda_z$, the effective width of the electron wave function along the growth direction \cite{Stano2018gfactor}. Here, $\lambda_z$ is given by the triangular confinement potential formed by the GaAs/AlGaAs heterostructure. We fit it from excited orbital state data and find $\lambda_z\approx6.5$\,nm similar for both devices \cite{Camenzind2018,Stano2017}, see Sec.~2 in Ref.~\cite{som}.

\begin{figure}[t]
	\centering
		\includegraphics[width=1\columnwidth]{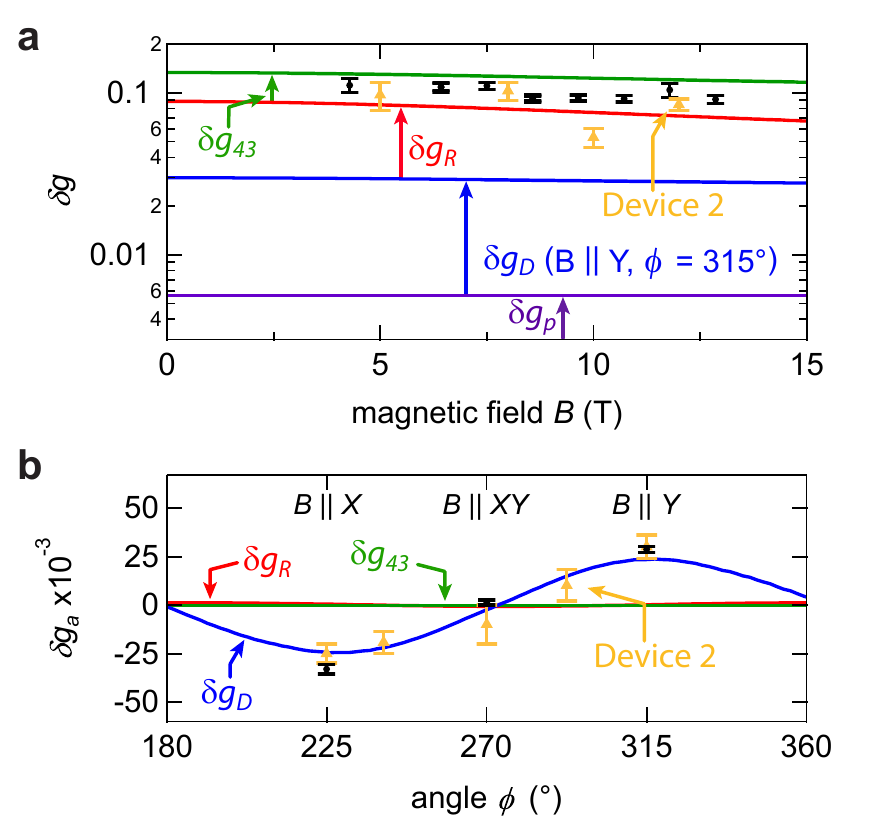} \vspace{-3mm}
	\caption{
(a)  Cumulative g-factor corrections $\delta g$ to $g_{\rm{bulk}}$ as labeled. The isotropic terms are due to penetration into the AlGaAs, $\delta g_p$, due to the $H_{43}$ term, $\delta g_{43}$, and the Rashba correction, $\delta g_{R}$. The Dresselhaus correction, $\delta g_D$, is anisotropic and given for a field along $Y$ ($\phi = 315^{\circ}$), the same direction for which the data is shown (black device 1, yellow device 2). Here, the g-factor corrections at the respective magnetic fields are directly obtained from the individual measured Zeeman splitting. The green curve shows the total theoretical g-factor correction for this field direction. In agreement with the model, the experiment barely shows any dependence on the magnetic field strength.
(b) The anisotropic corrections to the g-factor are dominated by $\delta g_D$ while $\delta g_{R}$ and $\delta g_{43}$ give insignificant anisotropic corrections.}
	\label{fig:gfactor:Fig3} \vspace{-4mm}
\end{figure}

We compare the experimental finding with the theoretical prediction for the magnetic field along $Y$ and the specific quantum dot confinement of device 1. We obtain $\delta g$, the g-factor correction from $g_{\rm{bulk}}$, from the measurement at each individual magnetic field by calculating $\delta g = |g_{\rm{bulk}}| - |\Delta/(\mu_{\rm{B}} B)|$. As seen in Fig.~\ref{fig:gfactor:Fig3}a, the data of the two devices are in agreement with each other within the error bars (apart from one outlier) and show a slight trend to decrease at large fields. Also, with most data points slightly below the green theory curve, it seems fairly clear that the theory overall predicts a somewhat larger correction than measured in experiment. While only one specific direction is plotted here, we find this discrepancy generally for the isotropic correction. The model predicts an average $\vert\bar{g}\vert = \vert g_{\rm{bulk}} + \delta g_{i} \vert \approx 0.33$ for an electron confined in such a quantum dot. The data presented in Fig.~\ref{fig:gfactor:Fig2}b suggests an isotropic correction to  $\vert \bar{g} \vert \approx 0.373$ for device 1, and $\vert \bar{g} \vert \approx 0.396$ for device 2. Thus, the theory calculates a stronger isotropic correction than seen in the experiment -- to be discussed later.

There are several predicted terms, as shown on Fig.~\ref{fig:gfactor:Fig3}a, which contribute to the isotropic correction $\delta g_{i}$. These terms originate either from the band structure or the heterostructure confinement, see Ref.~\cite{Stano2018gfactor}. The field direction only matters for the anisotropic Dresselhaus correction $\delta g_D$. From the theoretical calculations, we conclude that the isotropic correction is dominated by $\delta  g_R$, a correction due to intrinsic Rashba SOI, and by $\delta g_{43}$, a correction due to the generic SOI term $H_{43}$ \cite{Braun1985, Stano2018gfactor}. The well known Rashba SOI term originates from the structural inversion asymmetry in the GaAs/AlGaAs heterostructure, while $H_{43}$ is a bulk band structure term generated by a magnetic field. The next strongest isotropic term is the penetration correction $\delta g_p$ which arises from the overlap of the wave function with the AlGaAs bulk where $g_{\rm{AlGaAs}} = +0.4$ \cite{Hanson2003}. This term is negligible in our case but becomes substantial for smaller 2DEG widths ($\lambda_z \lesssim 4$\,nm). 


The anisotropic correction to the g-factor originates from the Dresselhaus SOI which is a consequence of bulk inversion asymmetry in the zinc blende crystal structure of GaAs. As seen in Fig.~\ref{fig:gfactor:Fig2}b, the largest correction to $g_{\rm{bulk}}$ is observed along $Y$. This is a strong indication that the Dresselhaus constant $\gamma_c $ is negative \cite{Stano2018gfactor,DettwilerPRX, Salis2001}, since a positive $\gamma_c$ would have the largest deviation from $g_{\rm{bulk}}$ in the $X$ direction. Our data suggests that $\delta g_a=0.030\pm0.002$ for device 1 and $0.025\pm0.003$ for device 2, which is close to the predicted $\delta g_a=0.024$. Further, for the relative correction to the g-factor, we find $\delta g_a / \vert \bar{g}\vert \approx 8.1 \pm0.5\%$ for device 1 and $\approx 6.3 \pm0.8\%$ for device 2, which is in good agreement with the model where this ratio is $\approx 7\%$.

We now discuss the possible origins of the discrepancy between theory and experiment in the isotropic correction. The first suspect is the lever arm: the accuracy of the mixing chamber temperature used here is not better than $5-10\%$, leading to a systematic uncertainty. Because we found that the lever arm is independent of both strength and direction of the field, the scaling of the g-factors along all directions would be equal. Thus, the accuracy of the g-factor measurement is limited by this systematic error to about $10\%$. For example, a smaller lever arm would result in a reduced $\vert \bar{g}\vert$, and hence a larger $\delta g_i$, closer to the model. Still, given the $\lesssim 10\%$ accuracy, this is not sufficient to reconcile the theory with the data from both devices. Despite the limited accuracy, the precision of the measurement originating from the statistical uncertainty is much better, around $1\%$, allowing us to compare e.g. g-factors along different directions with high resolution.

Another source of deviations could be that the constants used for the \textbf{k$\cdot$p} calculations in the model were off. From the data available, however, it is not possible to conclude which term leads to the overestimation of $\delta g_i$ when compared to the experiment.

Next, strain effects could be a source of the discrepancy: in the theory of Ref. \cite{Stano2018gfactor}, strain-induced SOI is not taken into account. Simplifications in the model of the heterointerface can also lead to a deviation from the observed g-factor: the model assumes an infinite linear slope of the triangular confinement potential and a step-like increase of the aluminum concentration at the AlGaAs/GaAs interface. In reality, the profile is different in both aspects: the linear slope levels off away from interface and there is a finite transition region from AlGaAs to GaAs. Perhaps most importantly here, the details of the interface on the atomic level can effectively induce additional spin-orbit interactions \cite{Rossler2002, Golub2004,Tanttu2018,Ferdous2018}.

Finally, we mention the possibility that one needs to go beyond \textbf{k$\cdot$p} theory to  fully account for our observations. For example, Ref. \cite{Bester2005} reports on self-assembled InGaAs/GaAs quantum dots which are so small and strongly strained that the structure inhomogeneities impose strong deviations from properties based on bulk crystal models. However, this scenario is rather improbable for our large and weakly strained (lattice matched) gated GaAs/AlGaAs dots.

\ \\

In summary, we find a clear g-factor anisotropy as well as an isotropic correction in two lateral, gate-defined quantum dots made on different GaAs/AlGaAs heterostructures. In one device, this ranges from $\vert g \vert = 0.344$ to $0.406$ depending on the direction of the applied magnetic field. We compare our findings to a recently proposed theory by \textit{Stano et al.} \cite{Stano2018gfactor} in which the g-factor corrections to the GaAs bulk value are divided into a leading isotropic and a weaker anisotropic part. While the measured isotropic corrections are weaker than predicted, our data for the anisotropic corrections are in good agreement with the theory. Here, the isotropic corrections arise from Rashba and $H_{43}$-type of spin-orbit interaction and the anisotropic correction originates from the Dresselhaus SOI. In silicon spin qubits, the anisotropy gives a change of the g-factor of order of one percent, dominated by surface roughness \cite{Tanttu2018,Ferdous2018}. In contrast, here, the measured anisotropy is more substantial due to the GaAs crystal lattice, i.e. the Dresselhaus SOI.

Our findings substantiate the relevant g-factor corrections in  GaAs spin qubits.
Here, the identification of the dominant g-factor correction terms could help to better understand the decoherence processes that originate from coupling to  charge noise, and in principle it could also be exploited for all-electrical spin manipulation.
Furthermore, this work represents a major step in probing band structure parameters using quantum dots. For example, extracting the \textbf{k$\cdot$p} parameters from bulk measurements is often complicated by effects of the electron-electron interaction. Here, since the dot is singly occupied, such effects do not enter.
This work shows that a quantum dot could be used as a well-controlled probe in experiments aiming at the microscopic parameters of the semiconductor. Similar experiments could be performed for samples with significantly different heterostructure confinements. From the dependence of the g-factor corrections on the width and symmetry of the heterostructure, the \textbf{k$\cdot$p} parameters could be obtained with a new level of confidence \cite{Stano2018gfactor}.

The data supporting this study are available in a Zenodo repository \cite{gfactorZenodo}.

\subsection{Acknowledgments}
We thank M. Steinacher and S. Martin for technical support. This work was supported by the Swiss Nanoscience Institute (SNI), NCCR QSIT and SPIN, Swiss NSF, ERC starting grant (DMZ), and the European Microkelvin Platform (EMP). PS acknowledges support from CREST JST (JPMJCR1675).

\section*{References}

%

\end{document}